\documentclass[a4paper,11pt]{article}
\usepackage{pos}

\usepackage{subfigure}

\def \jpsi {J/\psi}
\def \G {\gamma}
\def \GG {\gamma\gamma}

\def \kev {~\rm{keV}}

\title{Observation of the decay $\eta_c\to\gamma\gamma$ at BESIII}
\author*[a]{Zhi-Jun Li}
\author[a]{Zheng-Yun You}

\affiliation[a]{School of Physics, Sun Yat-sen University, Guangzhou 510275, China}

\emailAdd{lizhj37@mail2.sysu.edu.cn}
\emailAdd{youzhy5@mail.sysu.edu.cn}

\abstract{In the decays of $\eta_c\to\gamma\gamma$ and $J/\psi\to\gamma\eta_c$, there are discrepancies between the theoretical calculations and the PDG experimental values, referred to as the charmonium QCD puzzle. We observe the decay $\eta_c\to\gamma\gamma$ in $J/\psi\to\gamma\eta_c$ using $(2712.4\pm14.3)\times10^{6}$ $\psi(3686)$ events collected with the BESIII detector, and determine the product branching fraction $\mathcal{B}(J/\psi\to\gamma\eta_c)\times\mathcal{B}(\eta_c\to\gamma\gamma)=(5.23\pm0.26_{\rm{stat.}}\pm0.30_{\rm{syst.}})\times10^{-6}$. This measurement is consistent with the two most recent lattice QCD calculations, providing significant input toward resolving the charmonium QCD puzzle.}

\FullConference{The European Physical Society Conference on High Energy Physics (EPS-HEP2025)\\
7-11 July 2025\\
Marseille, France\\}

\begin{document}
\maketitle

\section{Introduction}
Charmonium systems serve as a valuable probe for investigating the nature of quantum chromodynamics (QCD), the fundamental theory governing strong interactions. However, in the decays $\eta_c \to \gamma\gamma$ and $J/\psi \to \gamma\eta_c$, discrepancies have been observed between theoretical predictions and the experimental values fitted by Particle Data Group (PDG), referred to as the charmonium QCD puzzle.

\begin{figure*}[htbp] \centering
	\setlength{\abovecaptionskip}{-1pt}
	\setlength{\belowcaptionskip}{10pt}

        \subfigure[]
        {\includegraphics[width=0.59\textwidth]{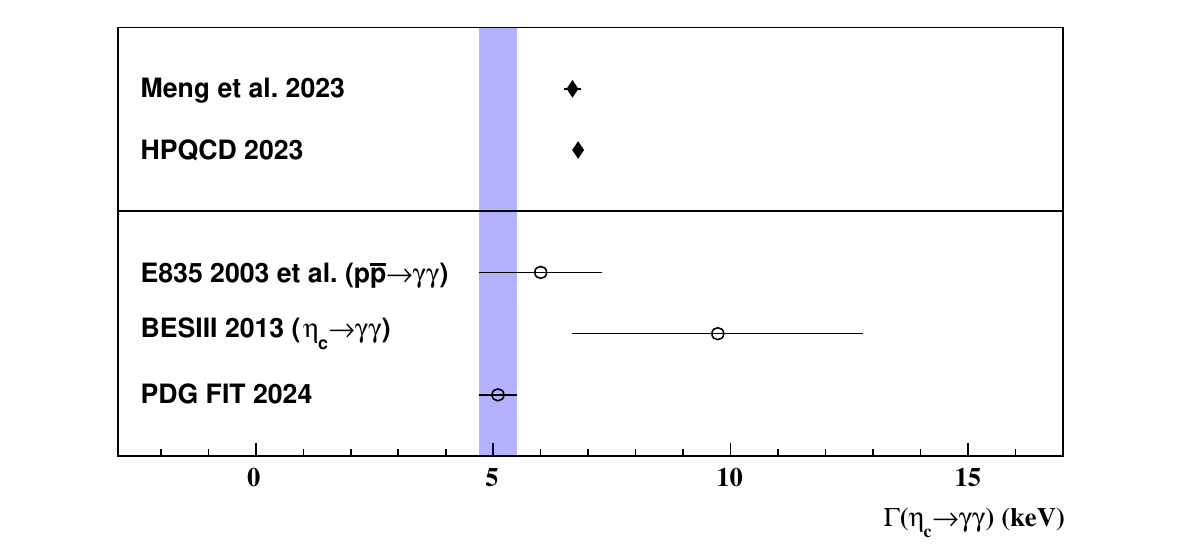}}
        \subfigure[]
        {\includegraphics[width=0.4\textwidth]{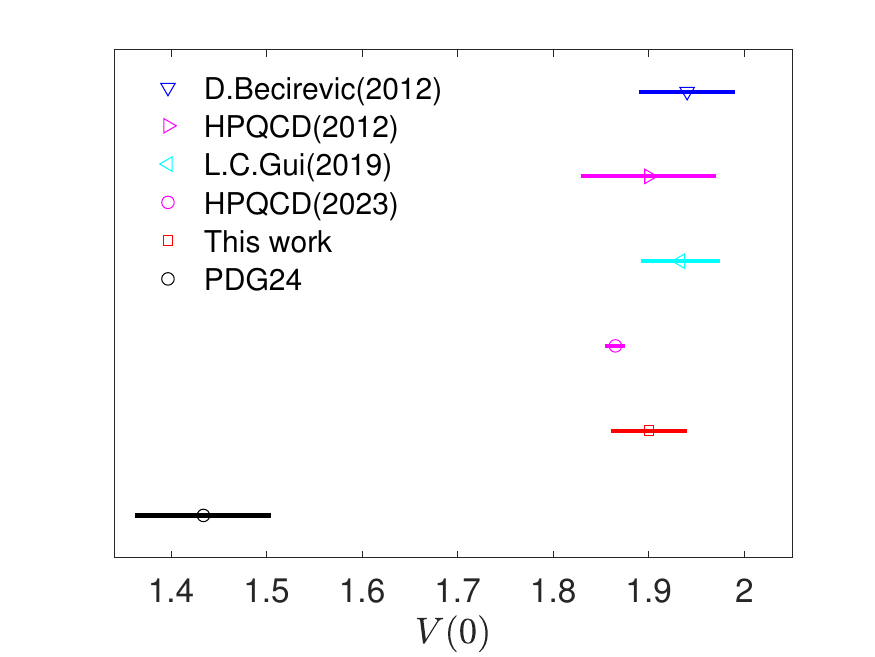}}\\
        
	\caption{
        (a) Comparison of the decay width $\Gamma(\eta_c \to \gamma\gamma)$ between LQCD calculations (top) and experimental measurements (bottom).
        (b) Comparison of $V(0)$ between LQCD calculations (colored points) and experimental values (black point)~\cite{Meng:2024axn}, where $|V(0)|^2 \propto \Gamma(J/\psi \to \gamma\eta_c)$.
        } 
	\label{fig:motivation}
\end{figure*}

Regarding the decay $\eta_c \to \gamma\gamma$, experimental evidence has been obtained through the process $J/\psi \to \gamma\eta_c$ by both the CLEO and BESIII collaborations, but with large uncertainties~\cite{CLEO:2008qfy,BESIII:2012lxx}. Additionally, $p\bar{p}$ annihilation experiments have measured the cross-section for $p\bar{p} \to \gamma\gamma$ at various energy points and observed a peak around the $\eta_c$ resonance~\cite{AnnecyLAPP:1987qkd,E760:1995rep,FermilabE835:2003ula}. Compared to these direct processes, measurements of the time-inverse process ($\gamma\gamma \to \eta_c$) are currently more precise~\cite{Belle:2018bry,Belle:2007qae,Belle:2012qqr,BaBar:2011gos,BaBar:2010siw,TASSO:1988utg}, and the world-average value of the decay width $\Gamma(\eta_c \to \gamma\gamma)$ from a global fit by the PDG is predominantly dominated by these $\gamma\gamma \to \eta_c$ measurements.
Theoretically, two recent Lattice QCD (LQCD) calculations~\cite{Meng:2021ecs,Colquhoun:2023zbc} deviate from the PDG world-average value by more than $3\sigma$, as illustrated in Figure~\ref{fig:motivation} (a). Other theoretical calculations exhibit varying degrees of deviation; some show greater discrepancies, while others remain consistent with the world-average value~\cite{Czarnecki:2001zc,Bodwin:2001pt,Brambilla:2018tyu,Yu:2019mce,CLQCD:2020njc,Dudek:2006ut,CLQCD:2016ugl,Chen:2016bpj,Li:2019ncs,Liu:2020qfz,Zhang:2021xvl,Feng:2017hlu}. These inconsistencies highlight an unresolved understanding of the $\eta_c \to \gamma\gamma$ decay within the framework of QCD.

For the decay $J/\psi \to \gamma\eta_c$, most theoretical calculations consistently show significant deviations from the world-average value fitted by PDG~\cite{Meng:2024axn,Colquhoun:2023zbc,Gui:2019dtm,Becirevic:2012dc,Donald:2012ga}, as illustrated in Figure~\ref{fig:motivation} (b) from Ref.~\cite{Meng:2024axn}.

%%%%%%%%%%%%%%%%%%%%%%%
\vspace{-0.0cm}
\vspace{-0.0cm}
%%%%%%%%%%%%%%%%%%%%%%%

\section{Results}
\begin{figure*}[htbp] \centering
	\setlength{\abovecaptionskip}{-1pt}
	\setlength{\belowcaptionskip}{10pt}

        \subfigure[]
        {\includegraphics[width=0.49\textwidth]{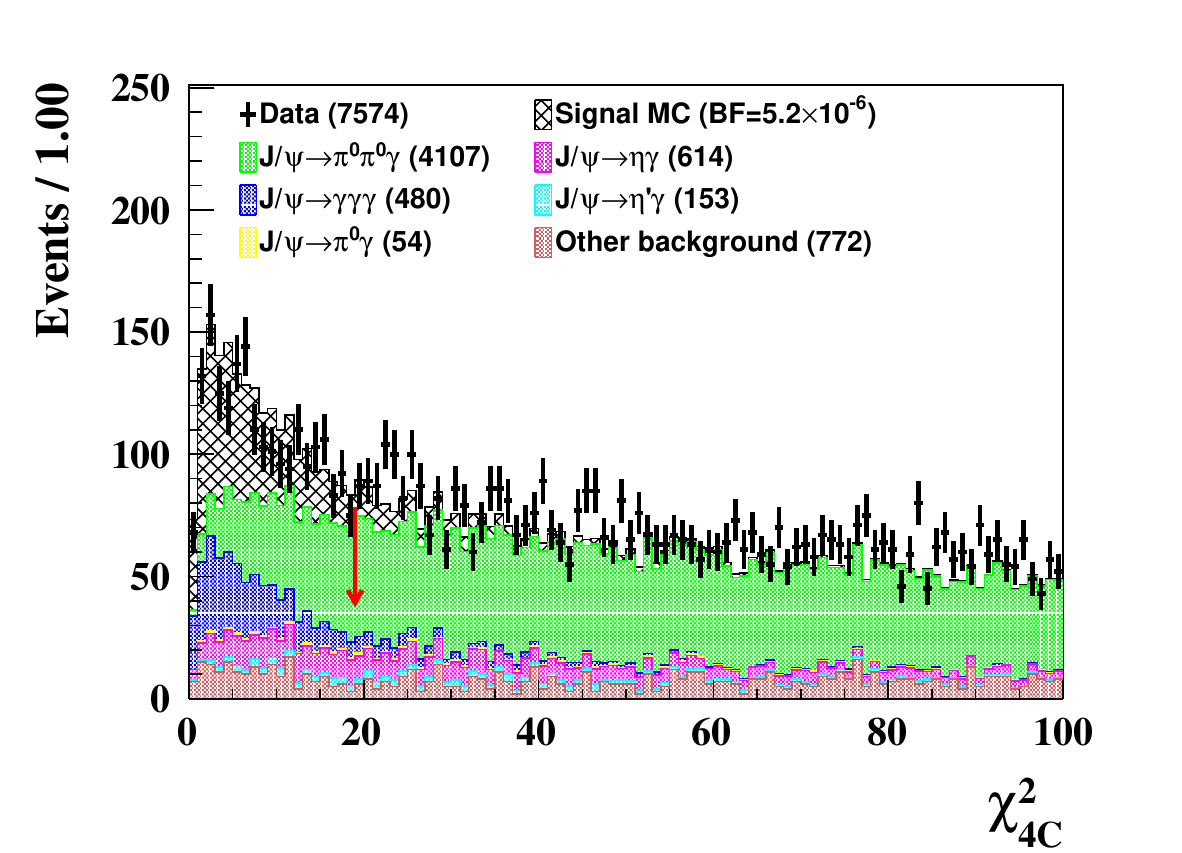}}
        \subfigure[]
        {\includegraphics[width=0.49\textwidth]{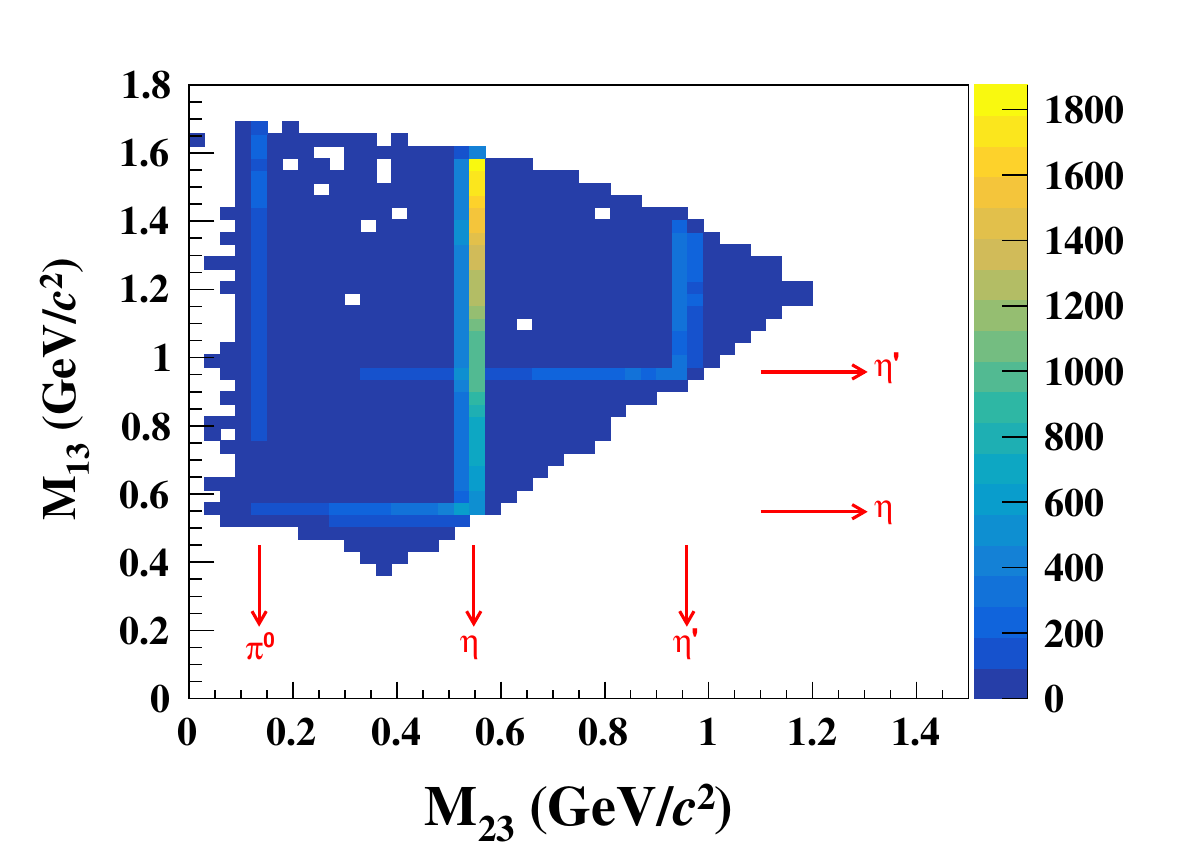}}\\
        \subfigure[]
        {\includegraphics[width=0.49\textwidth]{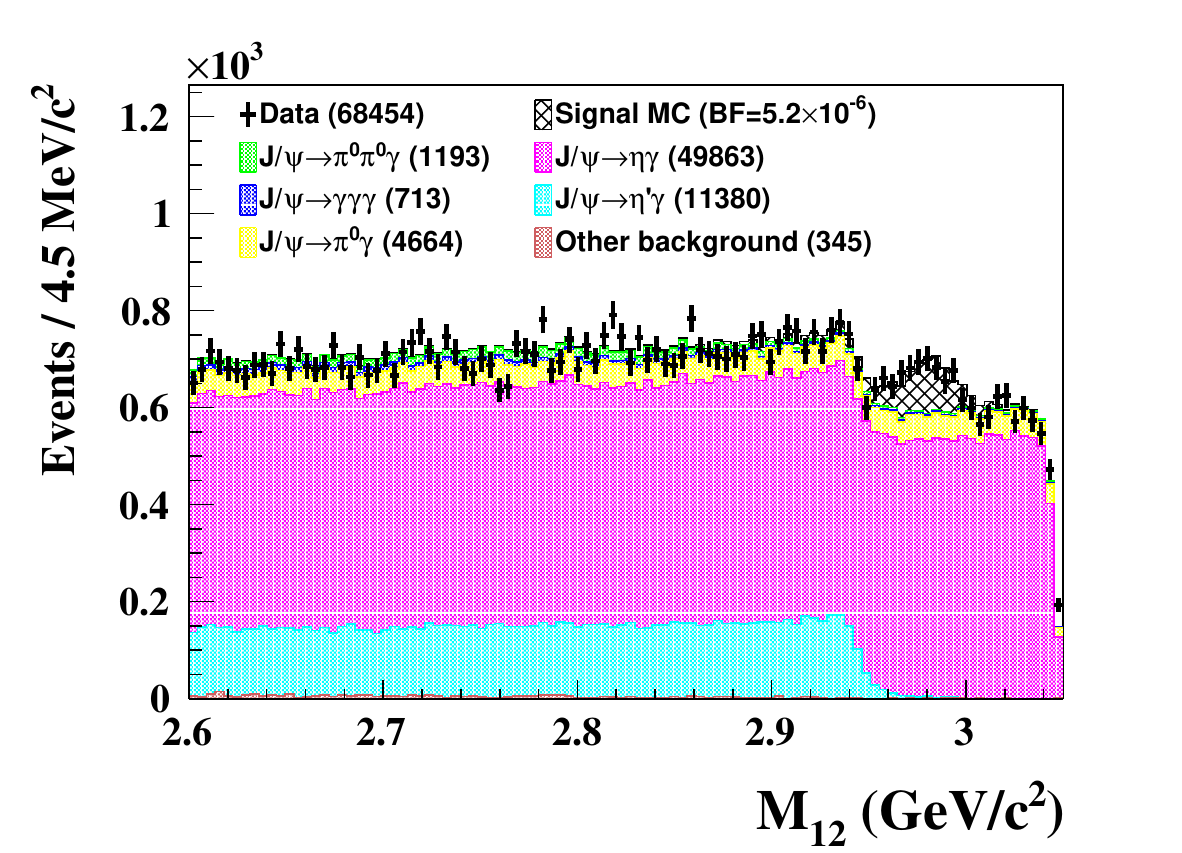}}
        \subfigure[]
        {\includegraphics[width=0.49\textwidth]{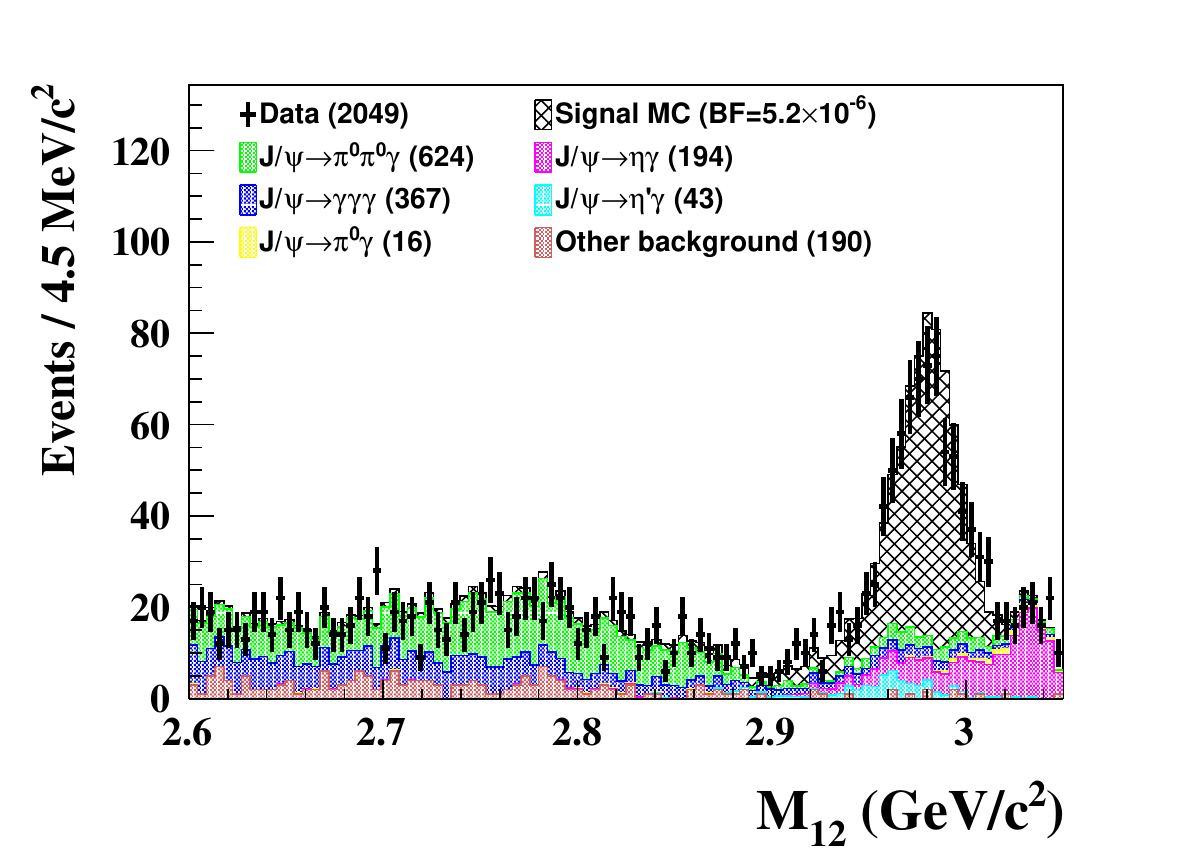}}\\
        
	\caption{
        (a) Distribution of $\chi^2_{4C}$. All selection criteria have been applied except for the $\chi^2_{4C}$ requirement.
        (b) Distribution of $M_{13} - M_{23}$ for the data samples.
        (c) Distribution of $M_{12}$ before applying the $M_{13} - M_{23}$ veto.
        (d) Distribution of $M_{12}$ after applying the $M_{13} - M_{23}$ veto.
        In (a), (c), and (d), the black points with error bars represent the observed data, while the colored histograms correspond to different background components as modeled by the MC simulation.
        } 
	\label{fig:cut}
\end{figure*}

\begin{figure*}[htbp] \centering
	\setlength{\abovecaptionskip}{-1pt}
	\setlength{\belowcaptionskip}{10pt}

        \subfigure[]
        {\includegraphics[width=0.32\textwidth]{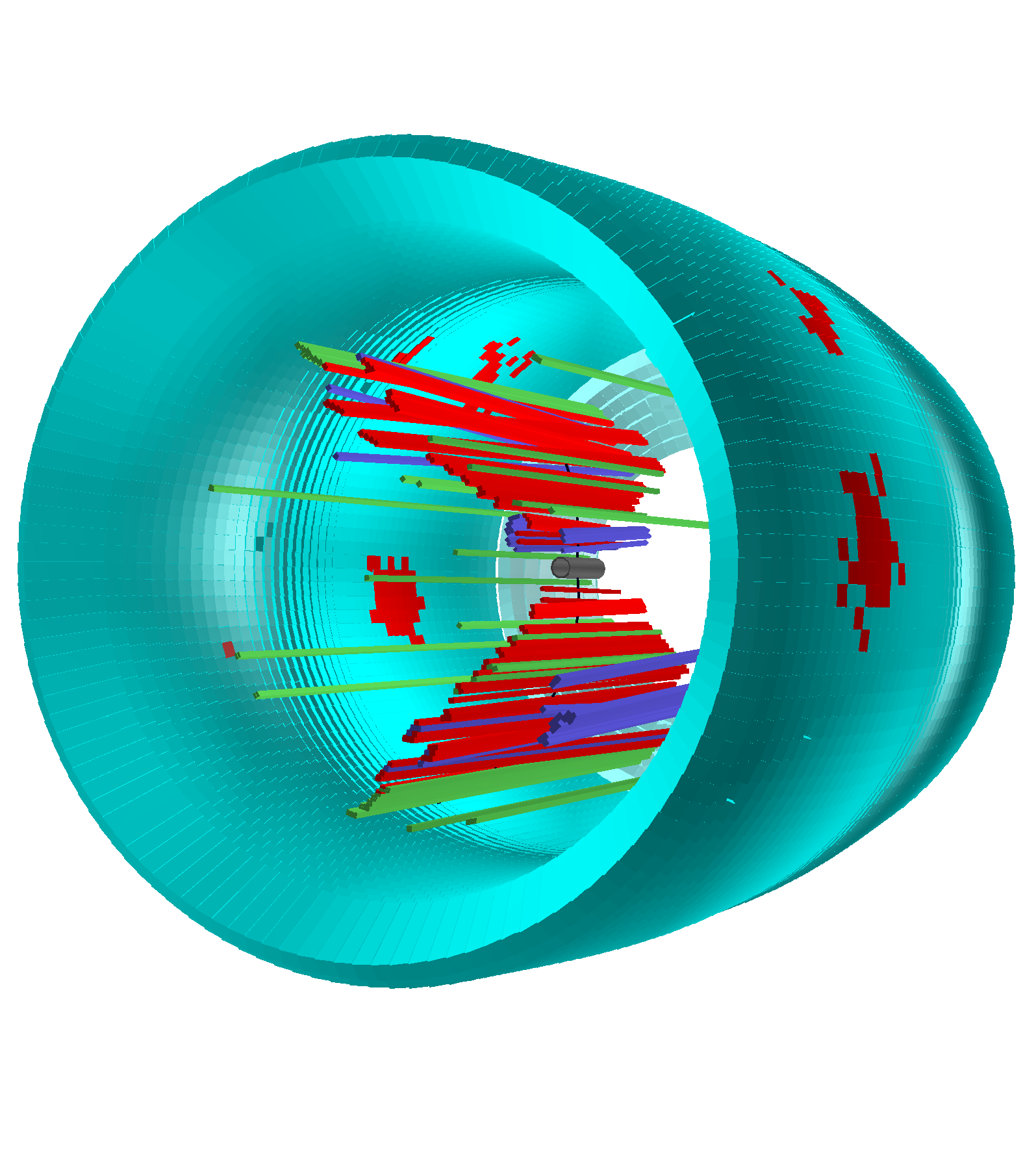}}
        \subfigure[]
        {\includegraphics[width=0.32\textwidth]{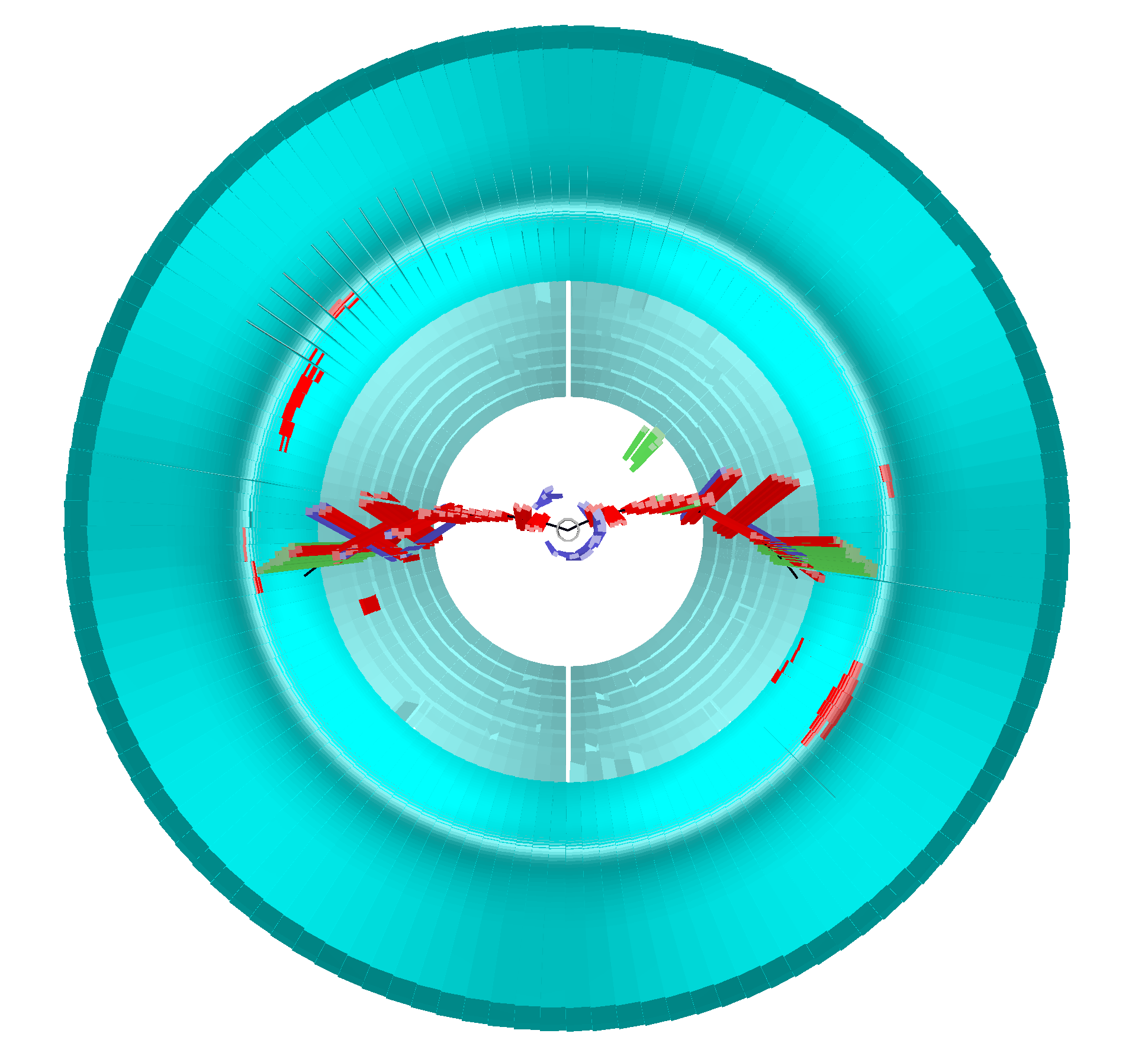}}
        \subfigure[]
        {\includegraphics[width=0.32\textwidth]{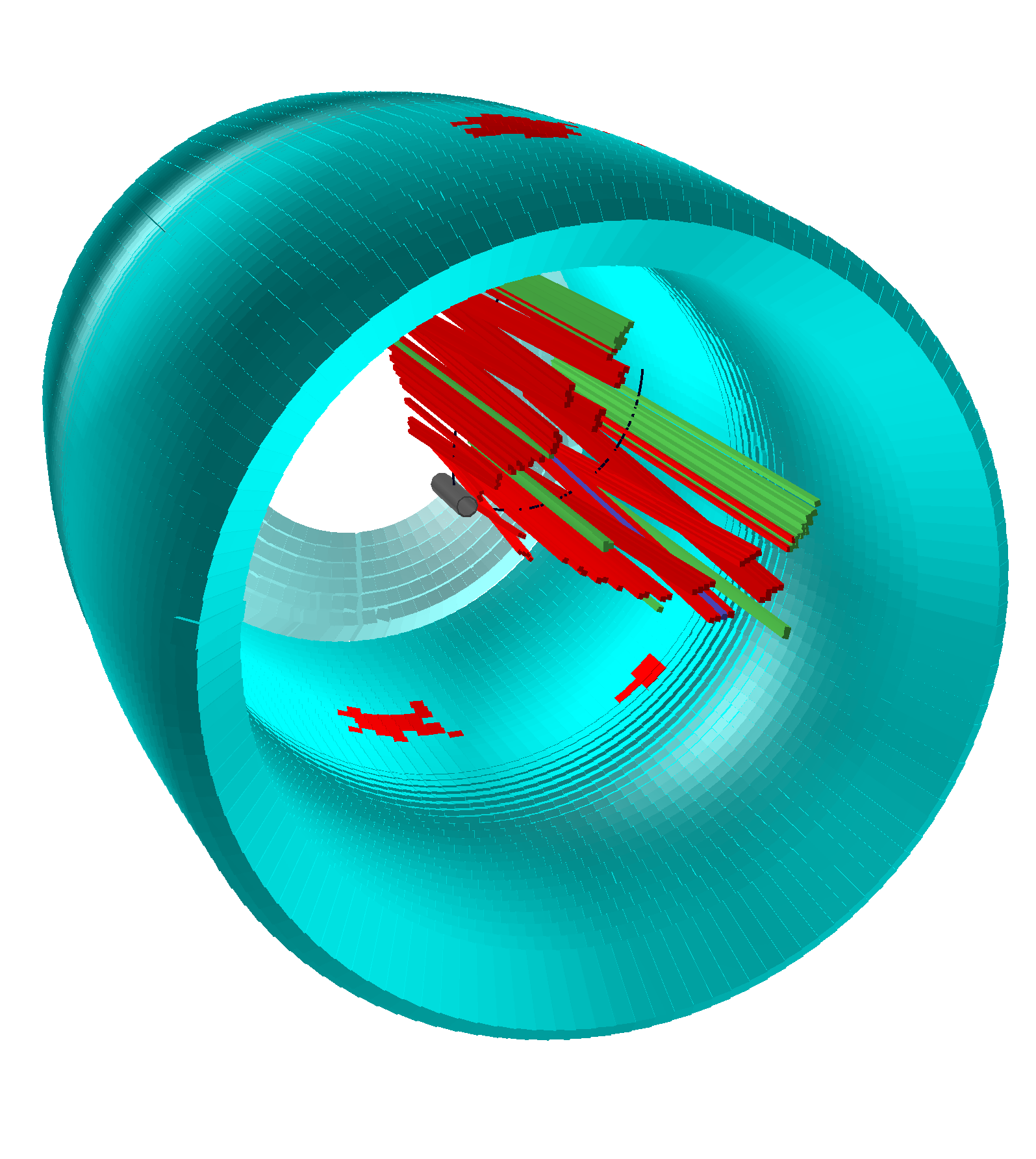}}\\
        
	\caption{
        Three signal candidates from real data in the event displays~\cite{Li:2024pox,Huang:2022wuo,Song:2025pnt}, collected on March 7, 2009 (a), December 31, 2011 (b), and June 28, 2021 (c), respectively. In these displays, the blue barrel represents the electromagnetic calorimeter (EMC) sub-detector, the red hits within the EMC correspond to showers from photons or charged tracks, and the internal hits denote charged tracks in the multilayer drift chamber (MDC).
        } 
	\label{fig:vis}
\end{figure*}

We present the measurement of $\eta_c \to \gamma\gamma$ via the decay chain $\psi(3686) \to \pi^+\pi^- J/\psi$ with $J/\psi \to \gamma\eta_c$ using $(2712.4 \pm 14.3) \times 10^{6}$ $\psi(3686)$ events~\cite{BESIII:2017tvm,BESIII:2024lks,Liao:2025lth} collected with the BESIII detector at the BEPCII collider. The final states consist of two charged pions and three photons, with no additional particles generated.
A four-constraint kinematic fit~\cite{Yan:2010zze} is performed, imposing the conservation of total four-momentum between the final state and the initial state. We require the chi-squared of the fit, $\chi^2_{\rm{4C}}$, to be less than 19 to ensure four-momentum conservation, as illustrated in Figure~\ref{fig:cut} (a). The three selected photons are labeled $\gamma_1$, $\gamma_2$, and $\gamma_3$, sorted by energy from highest to lowest. The $\eta_c$ signal is expected to peak in the invariant mass of $\gamma_1\gamma_2$ ($M_{12}$), while background processes involving $\pi^0$, $\eta$, or $\eta'$ typically peak in $M_{13}$ or $M_{23}$, as shown in Figure~\ref{fig:cut} (b). These backgrounds are suppressed by vetoing events in the $M_{13}$-$M_{23}$ spectrum. Further details of the event selection criteria can be found in Ref.~\cite{BESIII:2024rex}.
The $M_{12}$ distributions before and after applying the $M_{13}$-$M_{23}$ veto are shown in Figures~\ref{fig:cut} (c) and (d), respectively. After all selection criteria are applied, three signal candidates are presented as examples in Figure~\ref{fig:vis}.

%%%%%%%%%%%%%%%%%%%%%%%
\vspace{-0.0cm}
\vspace{-0.0cm}
%%%%%%%%%%%%%%%%%%%%%%%

%%%%%%%%%%%%%%%%%%%%%%%
\vspace{-0.0cm}
\vspace{-0.0cm}
%%%%%%%%%%%%%%%%%%%%%%%

\begin{figure*}[htbp] \centering
	\setlength{\abovecaptionskip}{-1pt}
	\setlength{\belowcaptionskip}{10pt}

        \subfigure[]
        {\includegraphics[width=0.49\textwidth]{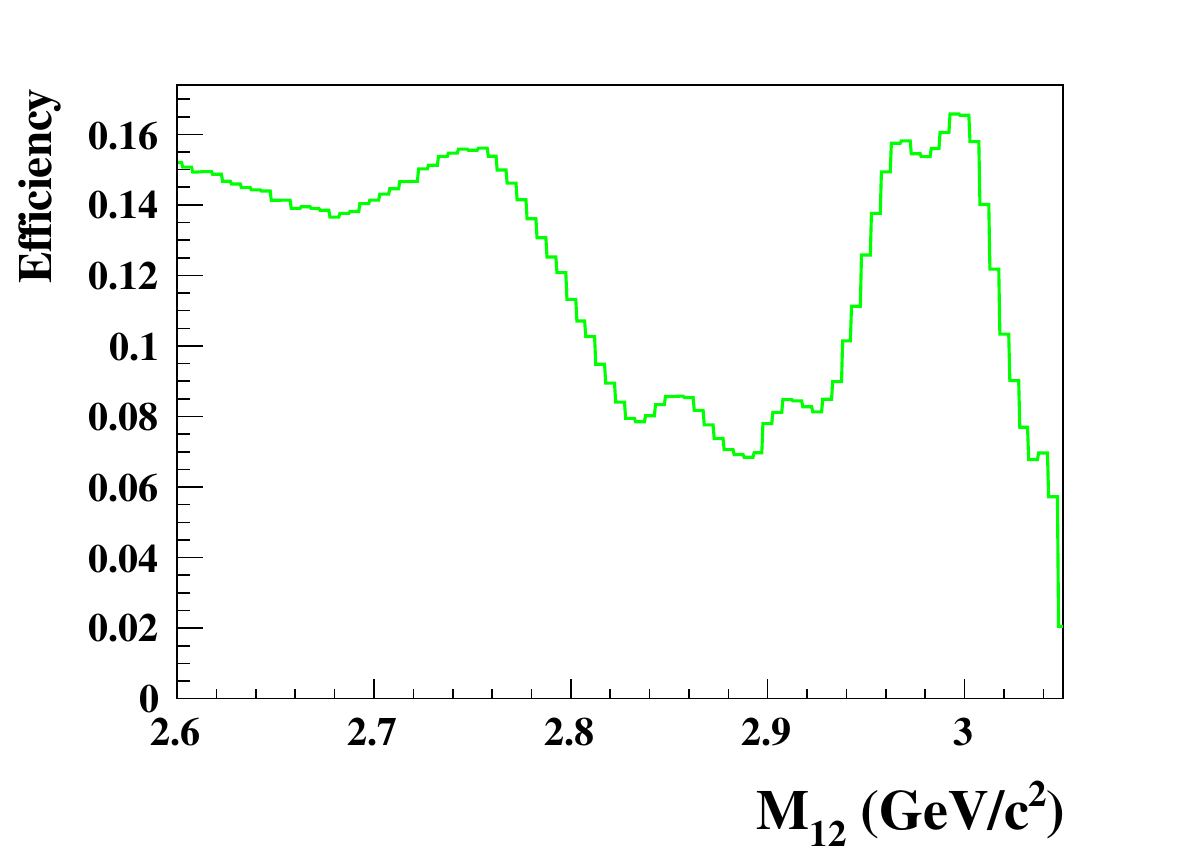}}
        \subfigure[]
        {\includegraphics[width=0.49\textwidth]{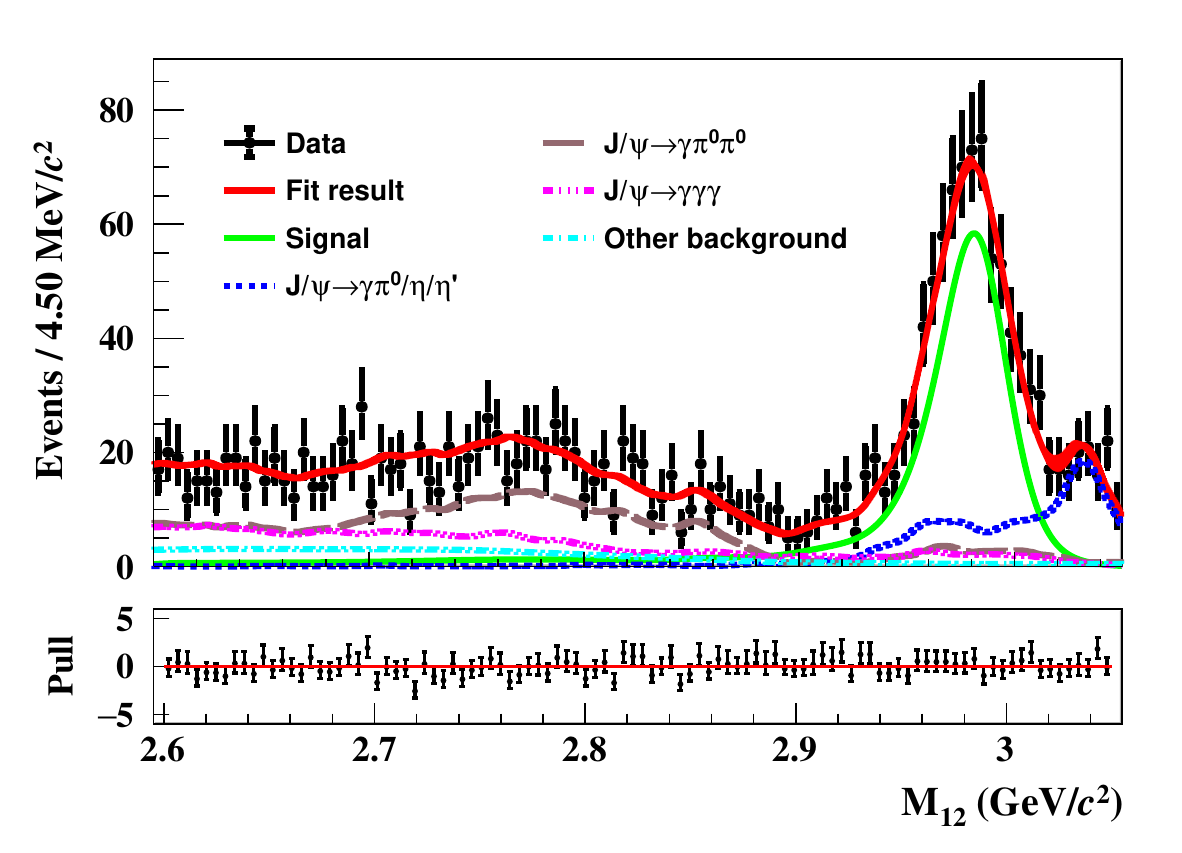}}\\
        
	\caption{
        (a) Mass-dependent efficiency derived from the signal MC simulation.
        (b) Fit to the $M_{12}$ distribution. The black points with error bars represent the data, the red line denotes the fit result, the green line corresponds to the signal, and the lines in other colors illustrate various background contributions.
        } 
	\label{fig:fit}
\end{figure*}

To extract the signal yield of $\eta_c \to \gamma\gamma$, we perform an unbinned extended maximum likelihood fit on the $M_{12}$ distribution. In this fit, the signal probability density function is modeled as
%\begin{eqnarray}
$\mathcal{PDF}(m) \sim \left[\epsilon(m) \times \mathcal{F}(m)\right] \otimes G(\mu, \sigma)$,
%\end{eqnarray}
where $\epsilon(m)$ represents the mass-dependent efficiency derived from signal MC simulations, as shown in Figure~\ref{fig:fit} (a). The function $G(\mu, \sigma)$ is a Gaussian with free parameters $\mu$ and $\sigma$ that account for the detector resolution, and $\mathcal{F}(m)$ denotes the lineshape of $\eta_c$ as described in Ref.~\cite{BESIII:2024rex}.
The background shapes are obtained from the MC samples using kernel density estimation~\cite{Poluektov:2014rxa}.
The fitting results are illustrated in Figure~\ref{fig:fit} (b), yielding a signal yield of $N_{\rm{sig}} = 677.7 \pm 33.5$. The product branching fraction (BF) $\mathcal{B}(J/\psi \to \gamma\eta_c) \times \mathcal{B}(\eta_c \to \gamma\gamma)$ is calculated to be $(5.23 \pm 0.26_{\rm{stat.}} \pm 0.30_{\rm{syst.}}) \times 10^{-6}$. Here, the relative systematic uncertainty is estimated to be $5.8\%$, and more details can be found in Ref.~\cite{BESIII:2024rex}.

%%%%%%%%%%%%%%%%%%%%%%%
\vspace{-0.0cm}
\vspace{-0.0cm}
%%%%%%%%%%%%%%%%%%%%%%%

\begin{figure*}[htbp] \centering
	\setlength{\abovecaptionskip}{-1pt}
	\setlength{\belowcaptionskip}{10pt}

        \subfigure[]
        {\includegraphics[width=0.49\textwidth]{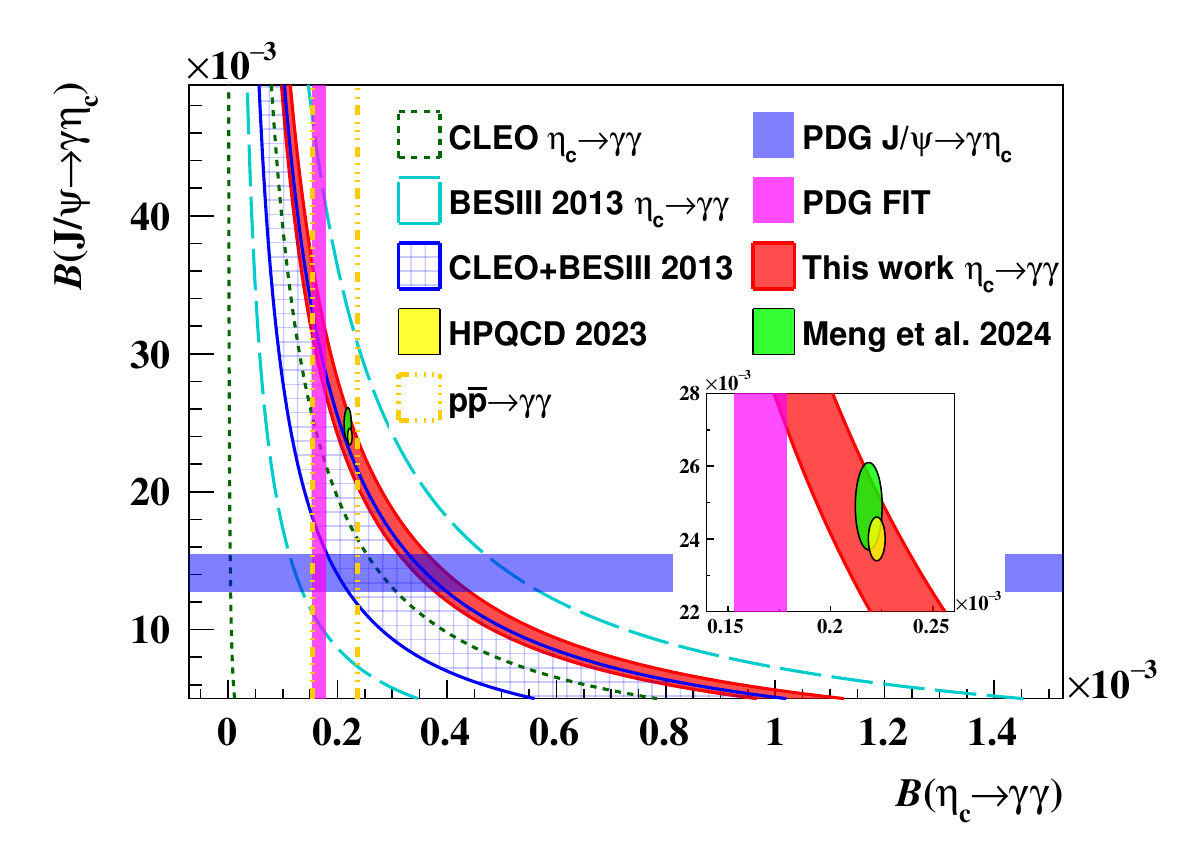}}
        \subfigure[]
        {\includegraphics[width=0.49\textwidth]{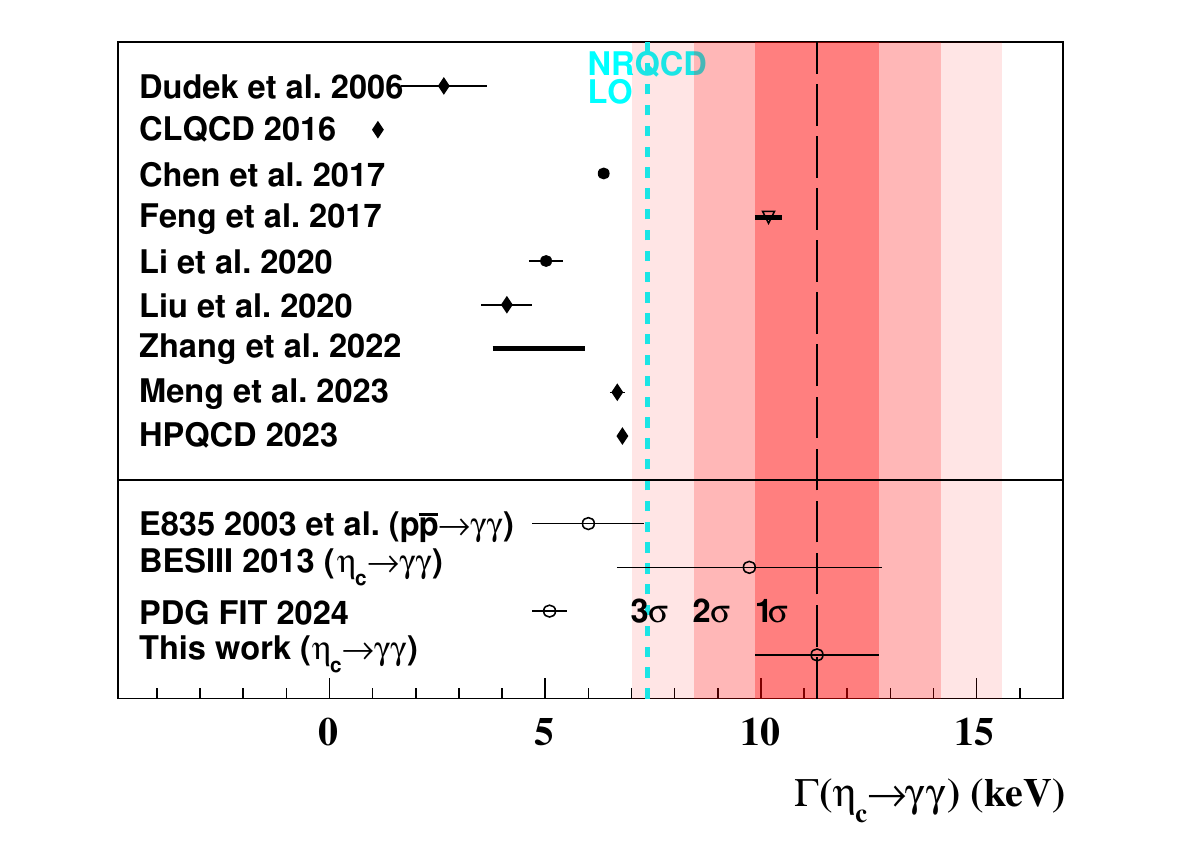}}\\
        
	\caption{
        (a) Comparison of $\mathcal{B}(\eta_c \to \gamma\gamma)$ versus $\mathcal{B}(J/\psi \to \gamma\eta_c)$ at the $1\sigma$ confidence level. The red-filled region represents our measurement, the yellow-filled and green-filled regions correspond to the LQCD calculations~\cite{Colquhoun:2023zbc,Meng:2021ecs,Meng:2024axn}. Other colored lines or filled regions indicate results from previous experimental measurements or PDG.
        %The green line represents the CLEO result~\cite{CLEO:2008qfy}, the light blue line represents the previous \mbox{BESIII} result~\cite{BESIII:2012lxx}, and the blue grid-filled region is the combined result of CLEO and BESIII. The purple-red filled region is the $\mathcal{B}(\eta_c\to\GG)$ result from the PDG~\cite{pdg:2024} and the blue full-filled region is the $\mathcal{B}(\jpsi\to\gamma\eta_c)$ result from the PDG~\cite{pdg:2024}. The orange line represents the average $p\bar{p}\to\gamma\gamma$ result from SPEC~\cite{AnnecyLAPP:1987qkd}, E760~\cite{E760:1995rep}, and E835~\cite{FermilabE835:2003ula} normalized to $\mathcal{B}(\eta_c\to p\bar{p})=(1.33\pm0.11)\times10^{-3}$.
        (b) Comparison of $\Gamma(\eta_c \to \gamma\gamma)$. In the plot labeled ``This work'', the black solid line represents the total uncertainty, including the reference uncertainties of $J/\psi \to \gamma\eta_c$ and $\Gamma_{\eta_c}$~\cite{pdg:2024}. The dark, dark-to-light, and light red-filled regions correspond to the $1\sigma$, $2\sigma$, and $3\sigma$ confidence levels, respectively. 
        %The LQCD calculations are indicated by rhomboid symbols, and the NNLO calculation of NRQCD is marked with an inverted triangle.
        } 
	\label{fig:comp}
\end{figure*}

 A comparison of our results with previous measurements~\cite{CLEO:2008qfy,BESIII:2012lxx}, the world-average values~\cite{pdg:2024}, and LQCD calculations from HPQCD~\cite{Colquhoun:2023zbc} and Meng et al.~\cite{Meng:2021ecs,Meng:2024axn} is shown in the $\mathcal{B}(\eta_c \to \gamma\gamma)$ versus $\mathcal{B}(J/\psi \to \gamma\eta_c)$ plane in Figure~\ref{fig:comp} (a).
In the plot, the value from the process $p\bar{p} \to \gamma\gamma$ is normalized using $\mathcal{B}(\eta_c \to p\bar{p}) = (1.33 \pm 0.11) \times 10^{-3}$~\cite{pdg:2024}.
We find that the world-average values of $\mathcal{B}(\eta_c \to \gamma\gamma)$ and $\mathcal{B}(J/\psi \to \gamma\eta_c)$~\cite{pdg:2024} do not simultaneously align with our measurement. Interestingly, the highly precise LQCD predictions from HPQCD, and those from Meng et al.~\cite{Meng:2021ecs,Meng:2024axn} both agree with our measurement, while the corresponding individual calculations of $\mathcal{B}(\eta_c \to \gamma\gamma)$ and $\mathcal{B}(J/\psi \to \gamma\eta_c)$ are inconsistent with the world-average values~\cite{pdg:2024}. No other theoretical calculations provide both $\mathcal{B}(\eta_c \to \gamma\gamma)$ and $\mathcal{B}(J/\psi \to \gamma\eta_c)$ simultaneously.

Using $\mathcal{B}(J/\psi \to \gamma\eta_c) = (1.41 \pm 0.14)\%$ and $\Gamma_{\eta_c} = (30.5 \pm 0.5)\,\text{MeV}$ from PDG~\cite{pdg:2024}, the decay width of $\eta_c\to\GG$ is determined to be $\Gamma(\eta_c\to\GG)=(11.30\pm0.56_{\rm{stat.}}\pm0.66_{\rm{syst.}}\pm1.14_{\rm{ref.}})\kev$,
with the first uncertainties statistical, the second systematic, and the third from $\mathcal{B}(\jpsi\to\G\eta_c)$ and $\Gamma_{\eta_c}$ used from the PDG~\cite{pdg:2024}.
Figure~\ref{fig:comp} (b) presents a comparison of the decay widths $\Gamma(\eta_c \to \gamma\gamma)$ from various theoretical calculations, experimental measurements, and the PDG world-average values. A notable discrepancy is observed between different experimental measurements and theoretical predictions.
\textbf{Note: Our determination of $\Gamma(\eta_c \to \gamma\gamma)$ depends on the chosen value of $\mathcal{B}(J/\psi \to \gamma\eta_c)$. Several theoretical calculations of $\mathcal{B}(J/\psi \to \gamma\eta_c)$ are significantly larger than the value fitted by PDG~\cite{Meng:2024axn,Colquhoun:2023zbc,Gui:2019dtm,Becirevic:2012dc,Donald:2012ga}.}

%%%%%%%%%%%%%%%%%%%%%%%
\vspace{-0.0cm}
\vspace{-0.0cm}
%%%%%%%%%%%%%%%%%%%%%%%

In summary, we observe the decay $\eta_c \to \gamma\gamma$ in $J/\psi \to \gamma\eta_c$, and the measured product BF is consistent with theoretical predictions, potentially alleviating the long-standing charmonium QCD puzzles. However, this study does not provide individual measurements of $\eta_c \to \gamma\gamma$ or $J/\psi \to \gamma\eta_c$, as determining one requires assuming the value of the other. Future experiments that independently measure either $\eta_c \to \gamma\gamma$ or $J/\psi \to \gamma\eta_c$ will be essential in further clarifying the current discrepancies.

\clearpage

\section*{Acknowledgements}
This work is supported by the National Key R\&D Program of China under Contracts Nos. 2023YFA1606000; National Natural Science Foundation of China~(Grant No. 12175321, 11975021).

% \begin{thebibliography}{99}
% \bibitem{...}
% ....

% \end{thebibliography}
\bibliographystyle{apsrev4-1}
\bibliography{mybib.bib}

\end{document}